\documentclass[twocolumn,10pt,fleqn]{article}
\usepackage{graphicx}
\usepackage{authblk}
\usepackage{hyperref}
\usepackage{cite}
\usepackage{subcaption}
\usepackage{float}
\usepackage{amsmath}
\usepackage{mathtools}
\usepackage{xcolor}
\usepackage{stfloats}

\title{Giant Enhancement of Phonon–Electron Coupling in Graphene under Femtosecond Laser Heating at Room Temperature}

\author[1,2]{Houssem Rezgui$^{*,\dagger}$}
\affil[1]{IMFT-CNRS, University of Toulouse, 7 Avenue du colonel Roche, Toulouse 31400, France}
\affil[2]{INL – International Iberian Nanotechnology Laboratory, Avenida Mestre José Veiga s/n, Braga 4715-330, Portugal}
\affil[$*$]{Corresponding author: houssem.rezgui@toulouse-inp.fr}
\affil[$\dagger$]{Submitted \today}
\date{} 

\begin{document}

\twocolumn[
  \begin{@twocolumnfalse}
    \maketitle
    \begin{abstract}
      In recent years, phonon-electron carrier dragging has emerged as an innovative approach for modulating energy transfer in low-dimensional systems. In this Letter, we explore the fundamental mechanisms of electron-phonon coupling and the role of thermal lag behavior in ultrafast heat transport. We present a theoretical investigation of non-equilibrium thermal dynamics in graphene under femtosecond laser excitation, emphasizing the role of phonon-branch-resolved electron–phonon coupling. This framework provides new insight into ultrafast energy transfer processes at femtosecond timescales and illustrates key deviations from the predictions of the classical two-temperature model (TTM), particularly in spatially localized heat transport. Our results show that a 190 fs laser pulse induces a strong non-equilibrium state, followed by momentum redistribution among the excited carriers. This is then followed by effective cooling of the carrier distribution on a 450 fs timescale through phonon emission.
    \end{abstract}
  \end{@twocolumnfalse}
]

The phonon gas kinetic theory describes heat conduction in solids by treating phonons, i.e., quantized lattice vibrations, as a gas of quasi-particles that transport thermal energy \cite{peierls1929,chen2005, Guo2015,Cao2007}. This approach models phonon dynamics using a Boltzmann-like transport equation, accounting for scattering processes such as phonon-phonon interactions \cite{allen2022,Guo2018}, phonon-electron coupling \cite{Zhou2020,Protik2020, Lu2018, Waldecker2016, Habibi2025}, and phonon-boundary scattering \cite{Nika2008,Balandin2011}, which influence thermal conduction. The Boltzmann transport equation (BTE) is often used to model phonon transport, accounting for their mean free path (MFP), group velocity, and relaxation time \cite{Johnson2013}. The BTE provides fundamental information on the thermal properties of materials, particularly in low-dimensional systems where the classical Fourier law becomes insufficient \cite{Cao2007,Guo2018, Chen2021}.

Electron-phonon coupling plays a crucial role in ultrafast thermal transport, particularly under extreme conditions induced by femtosecond laser heating \cite{Zhou2020,Habibi2025, Brorson1990, Rethfeld2002, Carpene2006, Miao2021,Lui2010,Shang2024,Ma2022}. When a femtosecond laser pulse interacts with a material, it deposits energy into the electronic subsystem, creating a highly non-equilibrium (n-eq) state \cite{Zhou2020, Miao2021}. Understanding these ultra-fast dynamics is crucial for optimizing heat dissipation in electronic devices. In particular, electrons transfer energy to the lattice via electron-phonon scattering, which typically occurs over a few picoseconds. This process generates phonons and leads to lattice heating, as shown in Figure 1. The rate of energy transfer is characterized by the electron-phonon coupling constant ($G$), which depends on the electronic density of the states of the material and the dispersion of the phonon \cite{Miao2021, Qiu1993}. Materials with strong electron-phonon coupling (e.g. Au, Cu, Ag) efficiently transfer energy to the lattice \cite{Lin2008, Qiu1993}, while weakly coupled materials (e.g. graphene) exhibit delayed phonon heating \cite{Habibi2025, Choi2021, Zobeiri2021, Mehew2024} as summarized in Table 1.

\begin{table}[ht]
\centering
\caption{Electron–phonon coupling constant \( G \) for graphene and metals.}
\label{tab:eph-coupling}
\begin{tabular}{|l|c|c|}
\hline
\text{Material} & \( G \) (W·m\(^{-3}\)·K\(^{-1}\)) & \text{Reference} \\
\hline
Graphene & \(1.59 - 3.1 \times 10^{15}\) & \cite{Zobeiri2021} \\
Al & \(3.62 \times 10^{17}\) & \cite{Miao2021} \\
Au & \(1.92 \times 10^{16}\) & \cite{Miao2021} \\
Ag & \(2.01 \times 10^{16}\) & \cite{Miao2021} \\
Cu & \(7.25 \times 10^{16}\) & \cite{Miao2021} \\
\hline
\end{tabular}
\end{table}

\begin{figure*}[t]
    \centering
    \includegraphics[width=0.7\textwidth]{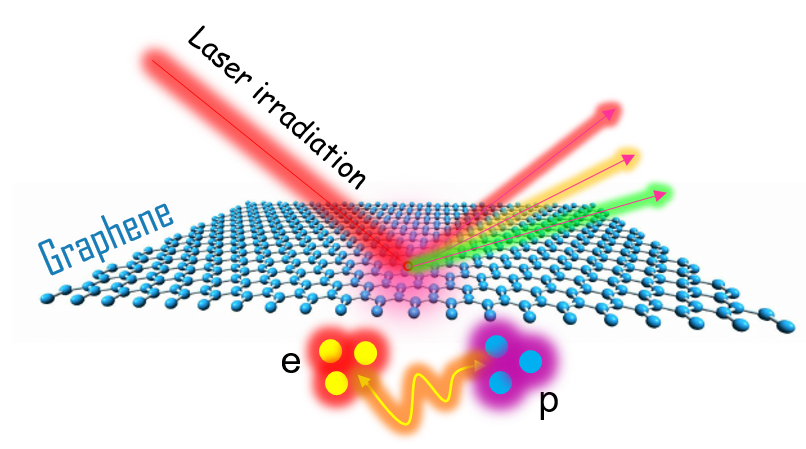}
    \caption{Ultrafast laser irradiation interacting with the graphene material.}
    \label{fig:fig1}
\end{figure*}

Graphitic materials are excellent candidates for studying electron-phonon interactions \cite{Habibi2025, Mehew2024}, phonon hydrodynamic transport \cite{Jeong2021, Lee2015}, second sound propagation \cite{Ding2022, Beardo2021, Rezgui2023}, and hot-electron cooling \cite{Mehew2024} due to their unique electronic and thermal properties. In graphene and other graphitic systems, the interaction between electrons and phonons enables a detailed exploration of energy-transfer mechanisms. The weak electron-phonon coupling allows for hydrodynamic electron flow, where electrons behave like a viscous fluid \cite{Sulpizio2019}. These properties make graphitic materials ideal for investigating hot-electron cooling dynamics, particularly through electron-phonon and phonon-phonon interactions. In high-quality material, hot-electron cooling occurs when phonons, generated by electron-phonon interactions, transfer their energy back to electrons, which then dissipate this energy, ultimately cooling the lattice \cite{Yu2023,Betz2012,Pogna2021,Breusing2009}. This cooling process was observed in the experiment on a short timescale, as short as 50 fs, in highly oriented pyrolytic graphite (HOPG) at an equilibrium temperature of 300 K \cite{Rohde2018}. It was found that an electron gas distribution forms simultaneously with the ultra-fast phonon-driven decay. However, if the emitted phonons have long lifetimes and do not rapidly decay into other phonon modes, they can re-absorb energy by interacting with the electrons \cite{Novko2019}. This feedback loop can slow down lattice cooling, as energy remains trapped within the electron-phonon system rather than dissipating efficiently into the lattice.

A widely used method to connect laser irradiation effects with a material's response is the two-temperature model (TTM), which describes the energy transfer between the electron and lattice systems \cite{Anisimov1974}. The model employs two differential equations to predict the temperature evolution of electrons and phonons, incorporating a coupling parameter associated with electron-phonon scattering. However, TTM is inadequate for capturing the ultrafast carrier dynamics due to the significant non-equilibrium effects that occur in extremely small systems \cite{Lu2018,Carpene2006}. Pogna et al. proposed a three-temperature model in which electrons, optical phonons, and acoustic phonons are treated as separate thermal reservoirs, each characterized by its own temperature. This framework captures the sequential energy transfer from electrons to optical phonons, followed by phonon–phonon scattering to acoustic phonons under ultrafast excitation conditions \cite{Pogna2021}. Qiu and Tien \cite{Qiu1993} developed the Hyperbolic Two-Step (HTS) Radiation Heating model by solving the Boltzmann equation to analyze the energy flux transported by electrons and the non-equilibrium state between electrons and the lattice during rapid heating events. However, both HTS and TTM do not account for different phonon branches, which can lead to inaccurate or misleading results. At extreme time and length scales, conventional Fourier heat conduction models are insufficient to describe the transport dynamics due to the thermal lag between the electron and phonon subsystems \cite{Cattaneo1958, Vernotte1958, Tzou1995, Tzou1997, Tzou2010}. To address these non-equilibrium effects, the Cattaneo–Vernotte (CV) model \cite{Cattaneo1958, Vernotte1958} offers a more precise framework by incorporating time lags: (1) the electron temperature lag, which accounts for the delay in energy transfer from the electrons to the lattice, and (2) the phonon temperature lag, which captures the delayed response of the lattice due to the finite speed of thermal wave propagation \cite{Ozisik1994, Tzou1995}. The CV model extends the standard TTM by integrating lagging effects \cite{Tzou1995,Tzou2010} and wave-like transport characteristics \cite{Ozisik1994,Guo2015,Beardo2021}, making it particularly effective for analyzing ultra-fast transport phenomena.

The aim of this letter is to investigate ultrafast thermal transport governed by electron-phonon coupling at room temperature in graphene. Inspired by the TTM and CV models, we report transient hot-electron cooling under femtosecond laser irradiation using an extended temperature model (ETM), which extends beyond the traditional two-temperature model and classical Fourier's law. In the present article, we propose a modification of the traditional TTM that accounts for lagging effects in different phonon branches.

Considering the delayed response time between the heat flux vector and the temperature gradient, the CV model is a modification of the classical heat conduction equation to account for thermal lag (i.e., finite-time delay in heat propagation) \cite{Ozisik1994, Tzou1995}. In classical heat conduction, the heat flux is assumed to propagate instantaneously, but this is not true, especially at high frequencies or in strong non-equilibrium thermodynamic conditions, where the thermal lag effect is significant. The CV model introduces a time-dependent thermal response to account for this lag effect. The CV or the so-called single-phase-lag (SPL) model can be represented for the phonon (\(p\)) and the hot-electron gas (\(e\)) as \cite{Tzou1995}:
\begin{equation}
\begin{aligned}
q_{p,i} + \tau_{p,i}\frac{\partial q_{p,i}}{\partial t} &= - \kappa_{p,i}\nabla T_{p,i}, \\
q_{e} + \tau_{e}\frac{\partial q_{e}}{\partial t} &= - \kappa_{e}\nabla T_{e},
\end{aligned}
\end{equation}
where \( q_{p} \) and \( q_{e} \) are the heat flux of phonon and hot electron, 
\( \tau_{q} \) and \( \tau_{e} \) are the lag phases of the heat fluxes, \( t \) is the time scale, \( \kappa_{p} \) and \( \kappa_{e} \) are the bulk thermal conductivities, \( T_{p} \) and \( T_{e} \) are the temperatures, and \( i \) is the index of phonon branches (LA, TA, and ZA) \cite{Lu2018}. The differential equation of the temperature evolution can be obtained by coupling Eq. (1) with a generalized TTM including electron supercollision \cite{Habibi2025,Laitinen2014}, as follows:
\begin{equation}
\begin{aligned}
C_{e}\frac{\partial T_{e}}{\partial t} &= - \nabla \cdot \overrightarrow{q_{e}} - \sum_{i} G_{ep,i}^{n}\left( T_{e} - T_{p,i} \right) \\
&- \sum_{i} G_{ep,i}^{SC}\left( T_{e} - T_{p,i} \right), \\
C_{p,i}\frac{\partial T_{p,i}}{\partial t} &= - \nabla \cdot \overrightarrow{q_{p,i}} + G_{ep,i}\left( T_{e} - T_{p,i} \right) \\
&+ G_{pp,i}\left( T_{lat} - T_{p,i} \right).
\end{aligned}
\end{equation}
Here, we define a phonon-electron coupling factor, \( G_{ep,i} = G_{ep,i}^{n} + G_{ep,i}^{SC} \), which allows for direct energy transfer between phonon branches. In graphene, energy transfer between the electron and the phonon occurs via normal, \( G_{ep}^{n} \) and supercollision scattering, \( G_{ep}^{SC} \), which denotes a non-linear coupling effect. Supercollision processes significantly enhance electron cooling in graphene by overcoming the momentum-conservation bottleneck of normal electron-phonon scattering. This allows thermal phonons to be exchanged more efficiently, boosting energy transfer between electrons and the lattice beyond what is possible in momentum-conserving interactions \cite{Song2015,Tielrooij2013}. The lattice temperature (\(T_{lat}\)), is considered to maintain energy conservation during transfer between the phonon branches \cite{Lu2018}:
\begin{equation}
\begin{aligned}
\sum_{i} G_{pp,i}\left( T_{lat} - T_{p,i} \right)=0
\end{aligned}
\end{equation}
The (\(p-p\)) scattering refers to the interaction between phonons, representing the coupling between each phonon branch and the scattering lattice reservoir. This process accounts for the exchange of energy between phonons within different branches and the surrounding lattice, ensuring that the energy redistribution among phonon modes is accurately modeled. The extended temperature model (ETM) can be derived by combining equations (1) and (2):
\begin{multline}
\tau_{e}\frac{\partial^{2}T_{e}}{\partial t^{2}} + \frac{\partial T_{e}}{\partial t}
= \frac{\kappa_{e}}{C_{e}}\nabla \nabla T_{e} \\
- \frac{G_{ep,i}^{n}}{C_{e}}(T_{e} - T_{p,i}) 
- \frac{G_{ep,i}^{SC}}{C_{e}}(T_{e} - T_{p,i}), \\
\tau_{p,i}\frac{\partial^{2}T_{p,i}}{\partial t^{2}} + \frac{\partial T_{p,i}}{\partial t}
= \frac{\kappa_{p,i}}{C_{p,i}}\nabla \nabla T_{p,i} \\
+ \frac{G_{ep,i}^{n}}{C_{p,i}}(T_{e} - T_{p,i})
+ \frac{G_{ep,i}^{SC}}{C_{p,i}}(T_{e} - T_{p,i}) \\
+ \frac{G_{pp,i}}{C_{p,i}}(T_{lat} - T_{p,i}).
\end{multline}

Equation (4) represents an extended version of the TTM, which accounts for the nonequilibrium state of the electron and phonon populations. We note that all input coefficients for different phonon branches are obtained using first-principles calculations \cite{Vallabhaneni2016}.
\begin{table*}[t]
    \centering
    \caption{Thermal Properties and Coupling Factors of Graphene at 297K}
    \small 
    \renewcommand{\arraystretch}{1.5}
    \begin{tabular}{lcccc}
        \hline
        Properties & Electron & LA & TA & ZA \\
        \hline
        $\kappa$ (W·m$^{-1}$·K$^{-1}$) & 50 & 863 & 237.9 & 2780 \\
        $C$ (MJ·m$^{-3}$·K$^{-1}$) & 0.00036 & 0.19 & 0.32 & 0.61 \\
        $G_{ep}^{n}$ (TW·m$^{-3}$·K$^{-1}$) & & 100 & 1 & 0 \\
        $G_{ep}^{SC}$ (MW·m$^{-3}$·K$^{-1}$) & & 93.9 & 413.6 & 0 \\
        $G_{pp}$ (TW·m$^{-3}$·K$^{-1}$) & & 2700 & 13000 & 1900 \\
        $\tau$ (ps) & 0.45 & 70.8 & 27.7 & 317 \\
        \hline
    \end{tabular}
    \label{tab:graphene_properties}
\end{table*}
Here, we present for the first time an extended temperature model (ETM) that goes beyond the classical TTM. We consider an initial room temperature condition (297 K) and a Gaussian laser excitation with different pulse durations at the top boundary. The single layer graphene (SLG) has lateral dimensions of 60 nm × 160 nm and the spot radius is about \( L_h = 35\,\text{nm} \). The electrons are heated to a maximum temperature of \( \Delta T_{\text{max}} = 1000\,\text{K} \). The probe beam is directed to the center of the hotspot to monitor the heat-pulse propagation. A single laser pulse, characterized by its pulse width (full width at half-maximum, FWHM duration), is introduced at the start of the simulation. The theoretical ETM predicts the transient thermal transport near the hot spot, while the other bondaries are assumed to be adiabatics. (See Supplementary Material S1). In a heat-pulse experiment setup, the adiabatic boundary is essential to observe ultrafast lattice cooling \cite{Jeong2021}. The pulse is modeled using a Gaussian temporal profile and is implemented within the finite element framework as a time-dependent boundary condition. We present a realistic design similar to the experimental configuration of transient thermoreflectance (TTR), which resembles the heat-pulse measurement geometry used in the second sound experiments \cite{Jeong2021, Ding2022,Qian2025}. Equation (4) is solved numerically using the finite element method within the COMSOL Multiphysics software \cite{Rezgui2024}. Here, our FEM solver is validated for several geometries against both theoretical and experimental data. A detailed description of the FEM solver, including its formulation and validation, is provided in our earlier works \cite{Rezgui2023, Rezgui2024}. The thermal properties and the (\(e-p\)) coupling factors, \( G_{ep} \) and \( G_{pp} \) are obtained from \cite{Lu2018} and listed in Table 2, while the relaxation time of \(\tau_{e}\) is obtained from \cite{Choi2018}. 
\begin{figure*}[t]
    \centering
    
    \begin{subfigure}[b]{0.49\textwidth}
        \centering
        \includegraphics[width=\textwidth]{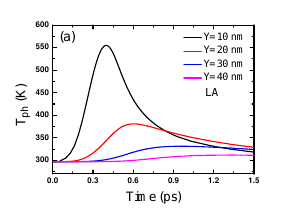}  
        \label{fig:fig2a}
    \end{subfigure}
    \hfill
    \begin{subfigure}[b]{0.49\textwidth}
        \centering
        \includegraphics[width=\textwidth]{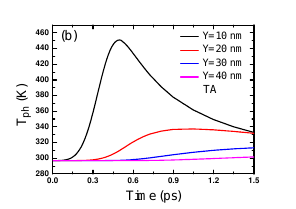}  
        \label{fig:fig2b}
    \end{subfigure}
    
    \vspace{-1em} 
    
    \begin{subfigure}[b]{0.49\textwidth}
        \centering
        \includegraphics[width=\textwidth]{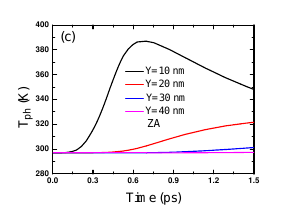}
        \label{fig:fig2c}
    \end{subfigure}
    \hfill
    \begin{subfigure}[b]{0.49\textwidth}
        \centering
        \includegraphics[width=\textwidth]{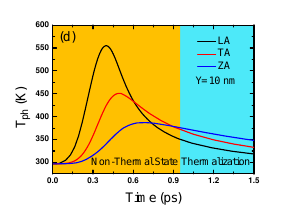}
        \label{fig:fig2d}
    \end{subfigure}

    \caption{Distribution of the temperature profiles according to the ETM for a pulse width of \(FWHM = 190\,\text{fs}\) at different \(Y\) positions. (a) \(LA\). (b) \(TA\). (c) \(ZA\). (d) Evolution of three phonon branches at \( Y = 10\,\text{nm} \).}
    \label{fig:fig2}
\end{figure*}

Figure 2 shows the transient temperature evolution of different branches of phonons in the center of the sample (\( X = 80\,\text{nm} \)) at various positions \( Y \). Figures 2(a), (b) and (c) show the temperature dynamics of the three phonon branches: longitudinal acoustic (LA), transverse acoustic (TA) and out-of-plane acoustic (ZA), highlighting the transient nonthermal state where phonon populations are not in thermal equilibrium. In Figure 2(d), the temperature of the LA phonons (\( T_{\text{p},LA} \)) increases more rapidly compared to the TA and ZA branches. In particular, \( T_{\text{p},LA} \) even exceeds its equilibrium value before eventually stabilizing and reaching thermal equilibrium with the other phonon modes. The lattice achieves thermal equilibrium, or carrier thermalization, at approximately 1.5 picoseconds. This thermal transition is strongly influenced by the phonon-phonon coupling constant and the laser pulse width. This behavior is characteristic of systems where energy transfer between electrons and phonons, as well as among different phonon modes, occurs on distinct timescales. The overshoot observed in \( T_{\text{p},LA} \) is due to its faster initial energy absorption, followed by a redistribution of energy to the TA and ZA branches over time.
\begin{figure*}[t]
    \centering
    
    \begin{subfigure}[b]{0.8\textwidth}
        \centering
        \includegraphics[width=\textwidth, keepaspectratio]{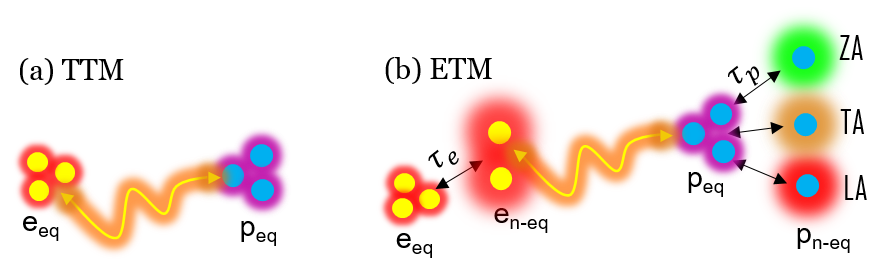}  
        \label{fig:fig3a}
    \end{subfigure}
    
    \vspace{-1em}

    \begin{subfigure}[b]{0.49\textwidth}
        \centering
        \includegraphics[width=\textwidth]{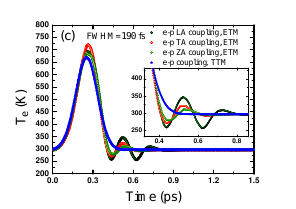}
        \label{fig:fig3c}
    \end{subfigure}
    \hfill
    \begin{subfigure}[b]{0.49\textwidth}
        \centering
        \includegraphics[width=\textwidth]{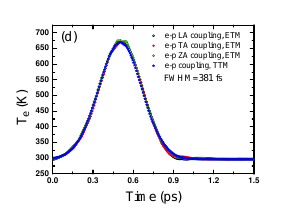}
        \label{fig:fig3d}
    \end{subfigure}
    
    \caption{Schematic representation of the (a) \(TTM\) and (b) \(ETM\). Evolution of the temperature of hotelectrons accroding to the ETM and TTM at \( Y = 20\,\text{nm} \) for different pulse width (c) \(FWHM = 190\,\text{fs}\) and (d) \(FWHM = 381\,\text{fs}\).}
    \label{fig:fig3}
\end{figure*}

Figures 3(a) and 3(b) provide a visual summary of the two-temperature and extended-temperature models, illustrating the nonequilibrium energy flow. In the ETM, the total energy of the lattice is determined by summing the contributions from all three phonon branches. This approach accounts for energy distribution across all lattice degrees of freedom, providing a comprehensive representation of the system's thermal state. Furthermore, the ETM incorporates the lagging response, a key characteristic of nonequilibrium transitions, which accounts for the delayed energy exchange between electrons and phonon branches within the lattice. Further comparison between TTM and ETM is provided in Supplementary Material S2.

Figures 3(c) and 3(d) illustrate the temperature evolution of hot electrons under long and short pulse widths using both the standard TTM and the ETM. A slight negative slope appears, indicating the activation of a cooling mechanism driven by electron-phonon interactions. Notably, under ultrafast laser irradiation, the ETM exhibits a wave-like behavior reminiscent of hydrodynamic second-sound transport. However, at room temperature, the momentum conservation of normal (\(N\))-scattering is not maintained, resulting in lattice cooling being dominated by the electron-phonon energy transfer process. The cooling time is on the order of a few femtoseconds and becomes significant for electron-phonon (\(e-p\)) coupling with LA phonons. The efficiency of hot-electron cooling depends on the presence of a short laser pulse width and factors such as electron and phonon lifetimes, which can provide additional cooling pathways. The electron temperature decreases further, becoming even lower than that of the LA phonons in ETM (See Figure S2 in Supplementary Material). In ultrafast processes, electrons thermalize faster than the phonon system can absorb their energy. This can result in electrons reaching a temperature lower than that of the phonon bath, especially under non-equilibrium conditions. The wavelike oscillations in the electron temperature become progressively weaker as the pulse width increases, illustrating a gradual transition from wave-dominated to diffusion-dominated cooling dynamics (See Figure S3 in Supplementary Material). The classical TTM model fails to capture the hot-electron cooling behavior under an ultrafast laser beam because of the absence of the thermal lag effect and electron interactions with different phonon branches in strongly non-equilibrium states. For long-pulse excitation, the TTM matches the ETM because the electron-phonon (\(e-p\)) coupling is weak, allowing sufficient time for the electron and phonon systems to reach thermal equilibrium. We further investigate the effect of excitation amplitude, defined by the peak electronic temperature increase \(\Delta T_{\text{max}}\), on the spatiotemporal heat transport dynamics. As \(\Delta T_{\text{max}}\) increases, we observe more pronounced wave-like transport behavior, indicative of more efficient carrier cooling. This behavior arises from stronger non-equilibrium conditions under higher excitation, which enhance electron–phonon coupling. The resulting dynamics indicate a regime where electron–phonon interactions govern energy redistribution, giving rise to wave-like features in the heat transport (See Figure S4 in Supplementary Material).
\begin{figure*}[t]
    \centering
    
    \begin{subfigure}[b]{0.49\textwidth}
        \centering
        \includegraphics[width=\textwidth, keepaspectratio]{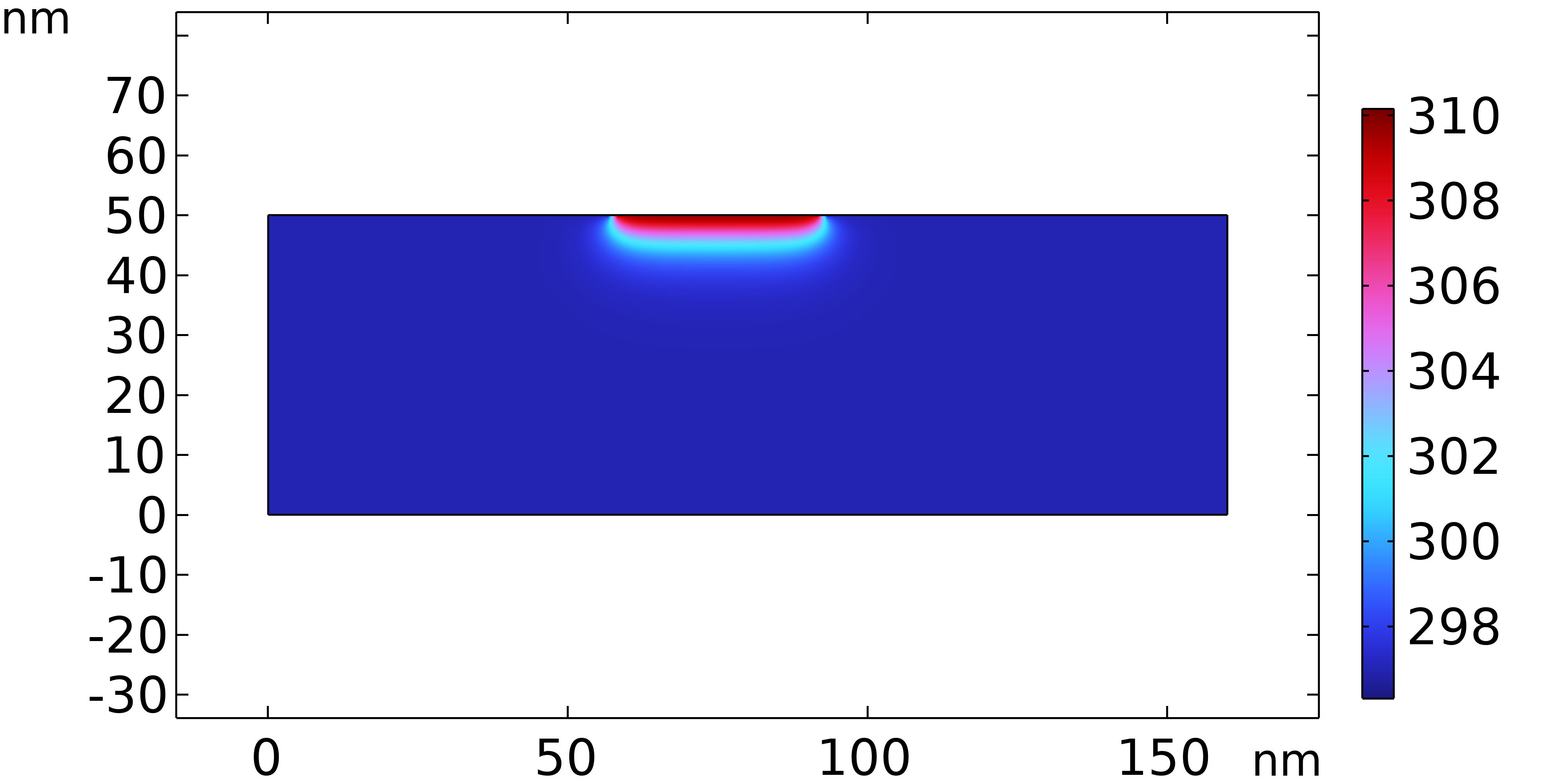}  
        \caption{10 fs}
        \label{fig:fig4a}
    \end{subfigure}
    \hfill
    \begin{subfigure}[b]{0.49\textwidth}
        \centering
        \includegraphics[width=\textwidth, keepaspectratio]{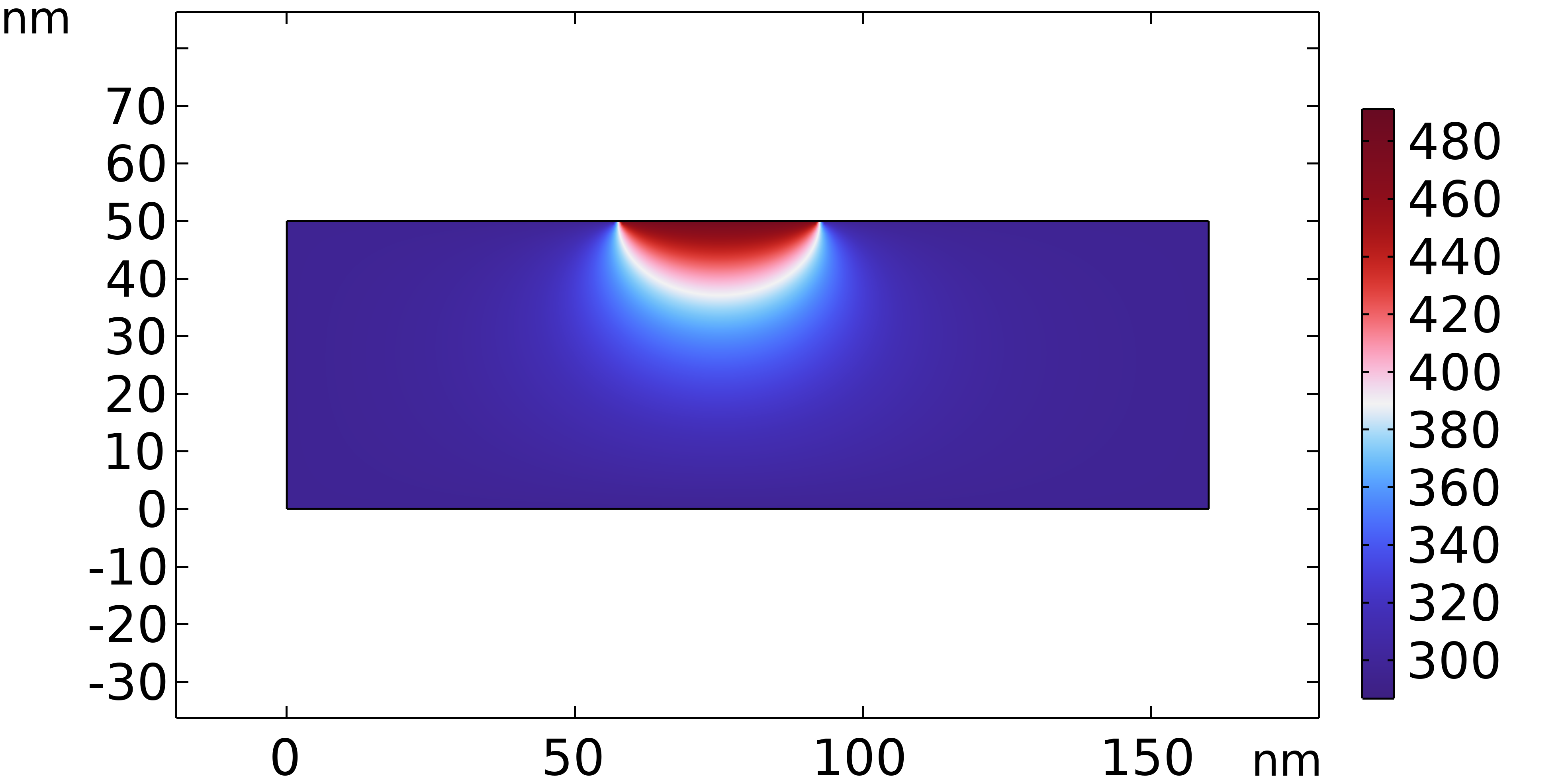}  
        \caption{100 fs}
        \label{fig:fig4b}
    \end{subfigure}
    
    \vspace{0.5cm}
    
    \begin{subfigure}[b]{0.49\textwidth}
        \centering
        \includegraphics[width=\textwidth, keepaspectratio]{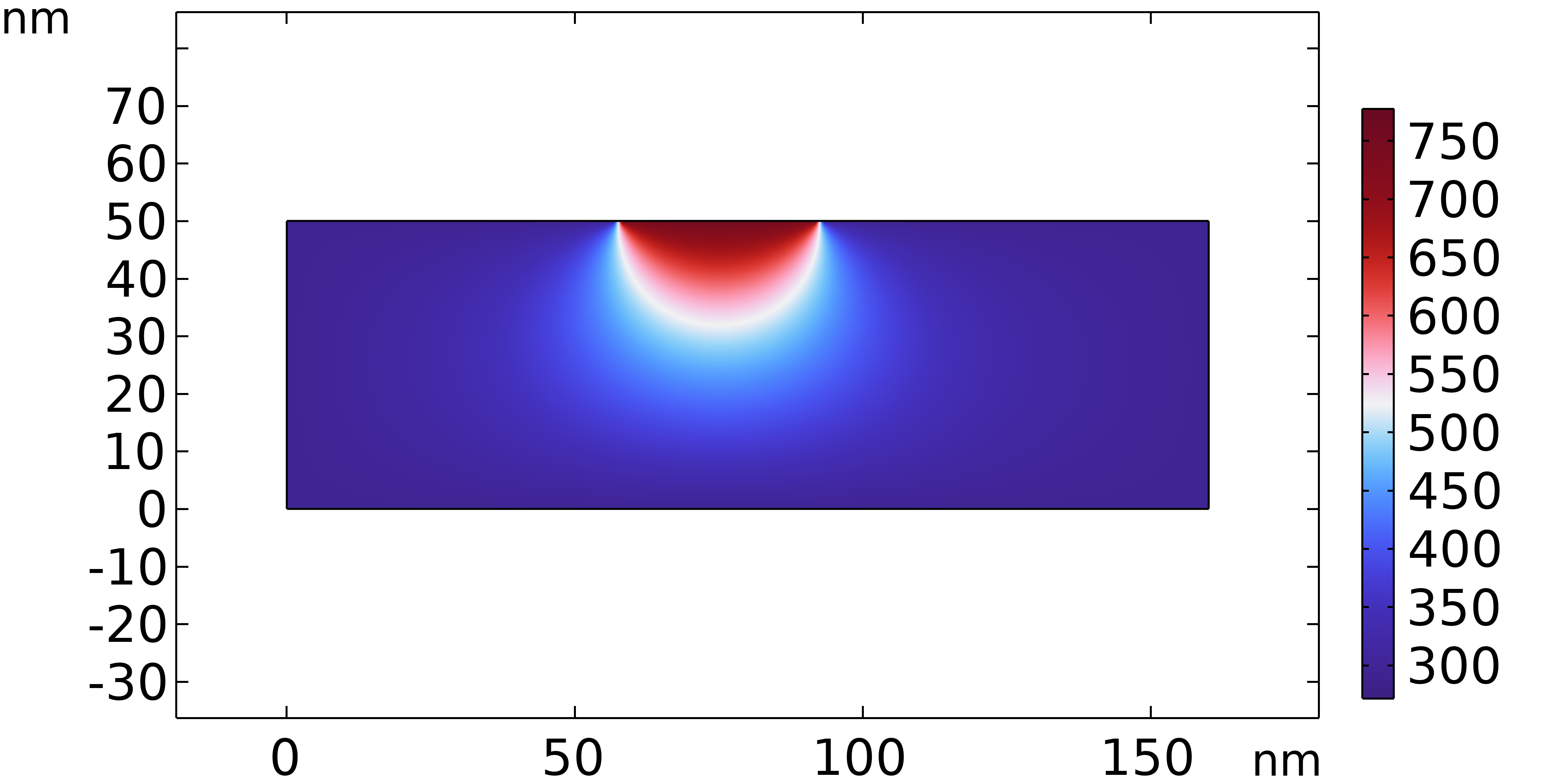}
        \caption{350 fs}
        \label{fig:fig4c}
    \end{subfigure}
    \hfill
    \begin{subfigure}[b]{0.49\textwidth}
        \centering
        \includegraphics[width=\textwidth, keepaspectratio]{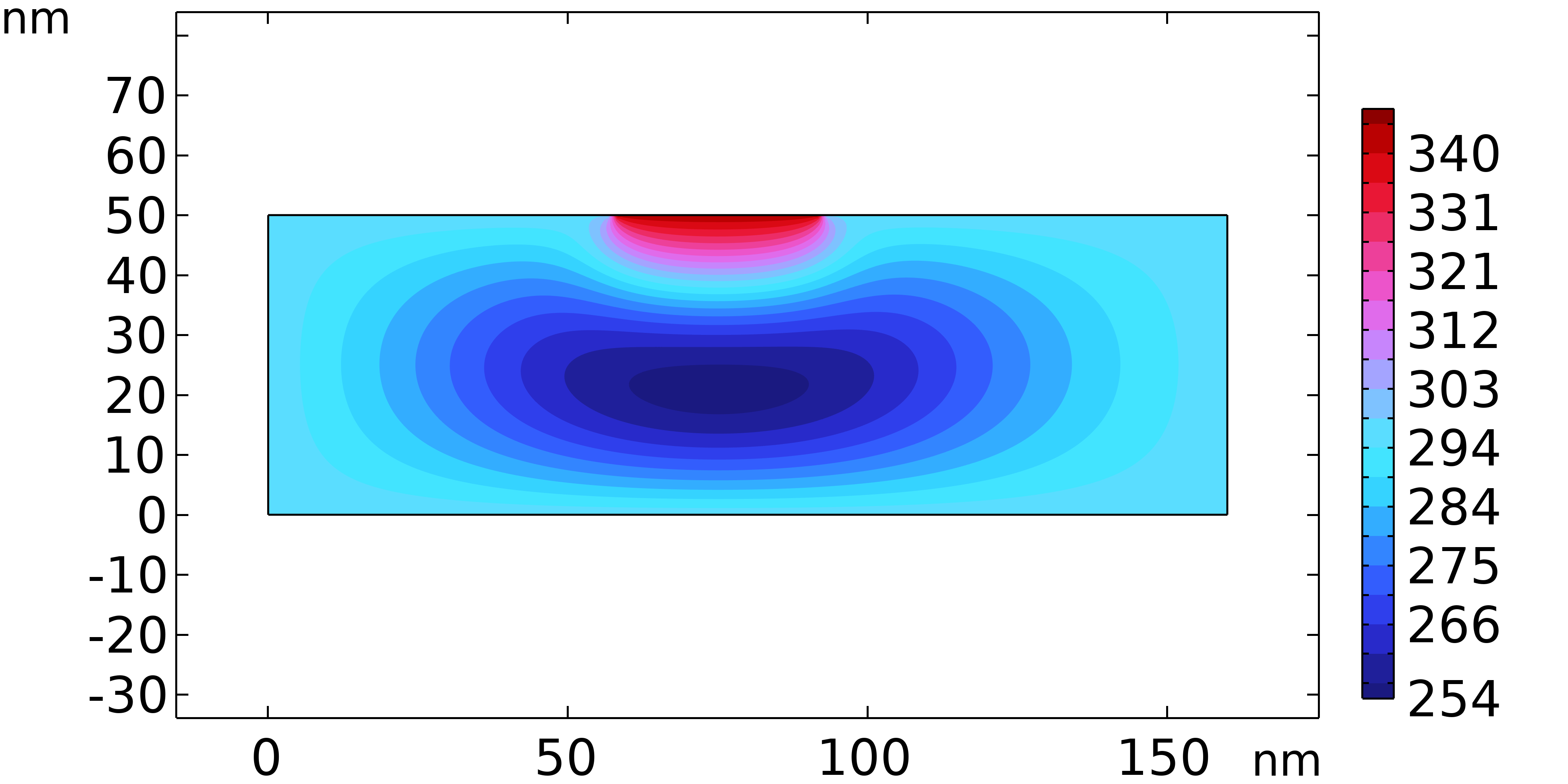}
        \caption{450 fs}
        \label{fig:fig4d}
    \end{subfigure}
    
    \caption{Transient nanoscale mapping of hot-electron temperature for a short pulse width \(FWHM = 190\,\text{fs}\).}
    \label{fig:fig4}
\end{figure*}

To finish, we demonstrate the mutual connection between the electron and phonon population during cooling dynamic. Figure 4 presents the surface and contour maps of the hot electron temperature for a short laser beam. In this analysis, we focus on the case of (\(e-p LA\)) coupling, as it exhibits the most significant cooling effect. As illustrated in Figures 4(a), 4(b), and 4(c), the electron temperature initially rises within the laser spot due to energy absorption and remains close to room temperature near the boundaries for \( t = 10, 100\,\text{fs}\) and \( t = 350\,\text{fs}\). This indicates that heat dissipation is relatively slow in the early stages. However, at \( t = 450\,\text{fs}\), an interesting phenomenon occurs: the electron temperature decreases significantly, becoming even lower than that of the LA phonons. A negative dip forms in the region heated by the laser pulse, causing the temperature to drop below its initial value. This behavior is characteristic of the electron gas, as it relaxes following excitation. It suggests that the temperature field exhibits wavelike behavior, with temperature oscillations, as illustrated in Figure 4(d). For electron-longitudinal acoustic phonon (\(e-p LA\)) interactions, the cooling time is approximately \( 180\,\text{fs}\) reaches a temperature \( T = 254\,\text{K}\). This relatively longer cooling time suggests that the interaction between electrons and longitudinal acoustic phonons is more efficient in preventing overheating and allows devices to operate at higher speeds. For electron-transverse acoustic and electron-out-of-plane acoustic phonon interactions, the cooling time is shorter, around \( 80\,\text{fs}\). As shown in Figure 3 (c), the lowest temperatures reached are \( T = 267.7\,\text{K}\) for (\(e-p TA\)) and \( T = 277\,\text{K}\) for (\(e-p ZA\)). This indicates less efficient energy transfer from electrons to these phonon modes, leading to shorter cooling times. This trend highlights the dominant role of electron-phonon interactions in the determination of thermal relaxation dynamics at ultrafast timescales, emphasizing their critical importance in device performance and heat management. A longer cooling time allows electrons to retain higher energy over a longer distance, improving carrier transport efficiency and enhancing the nanoscale thermal transport.

In summary, we investigated the ultrafast thermal dynamics in electron-phonon systems using both the standard two-temperature model (TTM) and the Extended Temperature Model (ETM). Our results reveal that phonon populations exhibit transient non-equilibrium states, where different phonon branches thermalize at distinct timescales. The longitudinal acoustic (LA) phonons play a dominant role in energy absorption and redistribution, leading to a characteristic overshoot in their temperature evolution before achieving thermal equilibrium. Furthermore, the ETM provides a more comprehensive understanding of ultrafast heat transport, capturing wave-like behaviors reminiscent of hydrodynamic second-sound transport. This effect is particularly pronounced under short-pulse laser excitation, where the electron temperature can briefly fall below that of the phonon bath as a result of the rapid energy transfer between electrons and phonons. This behavior, which the classical TTM fails to capture, highlights the significance of phonon-branch-resolved interactions in non-equilibrium thermal processes.

These insights have significant implications for nanoscale heat management and device performance. A deeper understanding of ultrafast electron-phonon interactions can aid in the design of high-speed electronic and optoelectronic devices by optimizing cooling pathways to prevent overheating. Future studies could extend this framework to explore material-dependent electron-phonon interactions, phonon hydrodynamic transport at room temperature, and the impact of external fields on ultrafast energy transfer.
\section*{Data availability statement}
All data supporting the findings of this study are available upon reasonable request of the authors.
\section*{Credit Author Statement}
\textbf{Houssem Rezgui}: Conceptualization, Formal analysis, Methodology, Software, Validation, Visualization, Writing – original draft \& editing.
\section*{Supplementary Material}
\noindent\hspace*{2em}The supplementary material provides additional details on the Extended Temperature Model (ETM), which expands on the traditional two-temperature model (TTM) and Fourier's law by accounting for the non-equilibrium nature of phonon systems. Specifically, it distinguishes between different phonon branches and incorporates thermal lag effects to better capture ultrafast thermal dynamics.


\end{document}


\maketitle

\newpage

\section{Ultrafast laser heating}

Femtosecond laser heating involves the use of ultra-short laser pulses, typically on the order of femtoseconds ($10^{-15}$ seconds), to rapidly deposit energy into a material. Due to the extremely short duration of the pulse, energy absorption occurs before significant heat diffusion, leading to highly localized heating and nonequilibrium conditions \cite{Lu2018}. In metals, for example, femtosecond laser heating primarily excites electrons, creating an electron-phonon non-equilibrium state before energy is transferred to the lattice \cite{Miao2021,Waldecker2016}. Here, this process is described by the extended temperature model (ETM), which accounts for the distinct temperature evolution of electrons and the lattice. Femtosecond laser heating is widely used in applications such as material processing \cite{Miao2021,Waldecker2016}, near-field thermal radiation \cite{Habibi2025,Becerril2023}, and studying ultrafast phase transitions at the nanoscale \cite{Novko2019,Yang2025}. The simalation is performed on a 160 × 50 nm of the SLG as shown in Figure S1. We use a single Gaussian pulse of 190 fs FWHM (short pulse) and 381 fs FWHM (long pulse).

\renewcommand{\thefigure}{S\arabic{figure}}
\begin{figure}[H]
    \centering
    
    \begin{subfigure}[b]{1\textwidth}
        \centering
        \includegraphics[width=\textwidth]{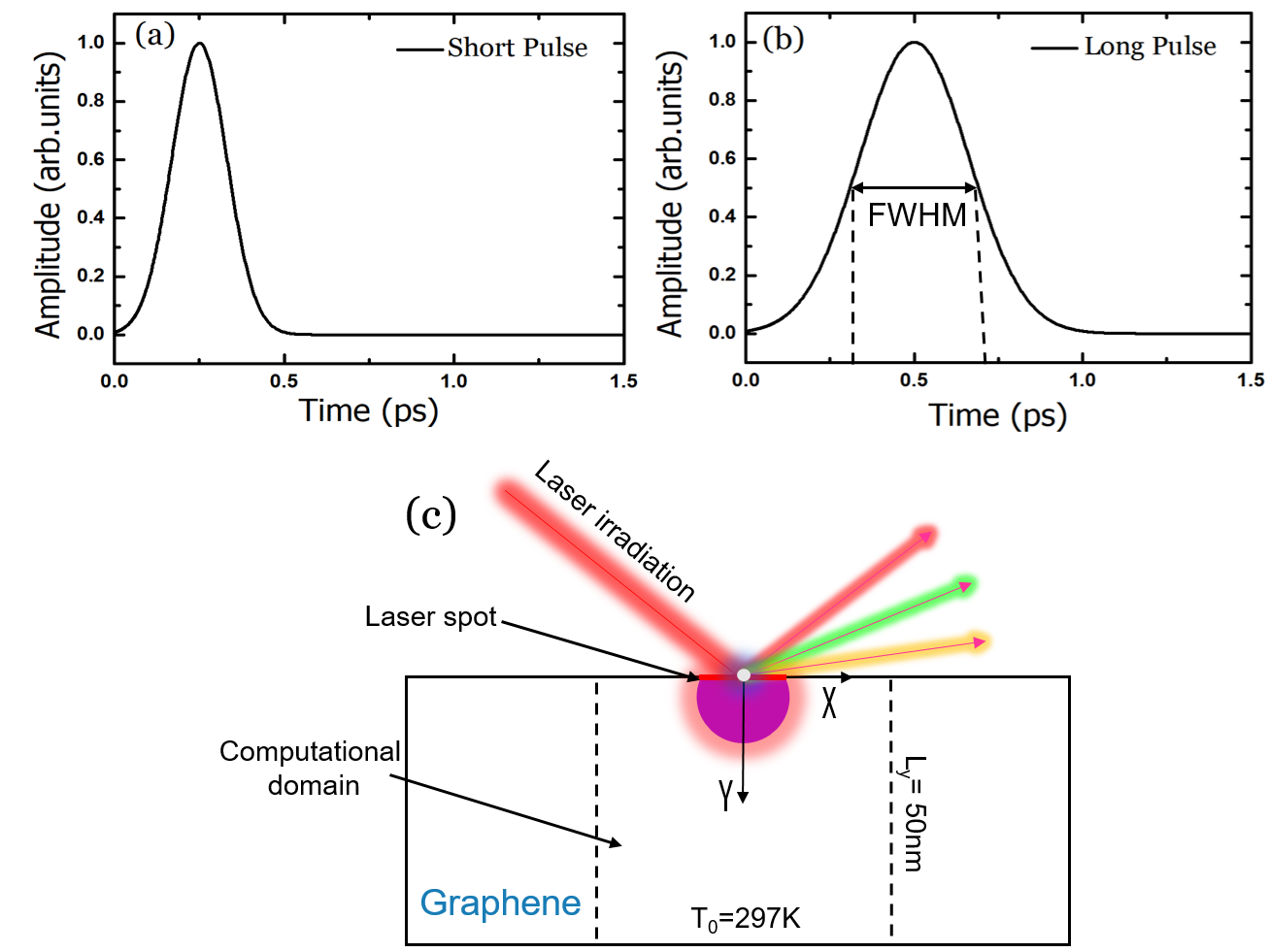}  
        \label{fig:fig2a}
    \end{subfigure}
    \hfill
    \begin{subfigure}[b]{0.49\textwidth}
    \end{subfigure}

    \caption{(a) Short pulse width of \(FWHM = 190\,\text{fs}\). (b) Long pulse width of \(FWHM = 381\,\text{fs}\). (c) An Illustration depicting laser heating in graphene.}
    \label{fig:fig1}
\end{figure}

\section{Beyond the two-temperature model}

The most common method to describe phonon transport is to use the Bose-Einstein (BE) distribution to characterize non-equilibrium quantum statistics within the framework of the Boltzmann equation. Non-equilibrium conditions emerge when a material experiences a rapid change or perturbation, causing it to deviate from its thermal equilibrium state, such as a high electric field, excitation by femtosecond laser pulses, or thermal radiation. However, in transient heat conduction processes, Fourier’s law implies an unphysical result of infinitely fast heat propagation, as the corresponding heat conduction equation is parabolic \cite{Tzou1995,Guo2018,Chen2021}. This physical limitation has led to numerous efforts to refine Fourier’s model. Cattaneo, followed by Vernotte introduced a modified heat conduction model known as the Cattaneo-Vernotte (CV) equation \cite{Cattaneo1958,Vernotte1958}. Unlike Fourier’s law, the CV equation is hyperbolic due to an additional term that involves the time derivative of the heat flux, ensuring a finite heat propagation speed. The CV equation is a solution to the phonon Boltzmann Transport Equation (BTE) obtained using the maximum entropy moment method to derive macroscopic equations for non-Fourier heat transport at room temperature \cite{Tzou1995,Guo2018}:

\begin{equation}
\begin{aligned}
q + \tau\frac{\partial q}{\partial t} &= - \kappa\nabla T \\
\end{aligned}
\end{equation}

where \( q \) is the heat flux of the phonon, \( \tau \) is the lag phase of the heat flux (i.e., the phonon relaxation time), and \( \kappa \) is the bulk thermal conductivity. However, in real situations involving an ultrafast pulse, one of the most intriguing challenges in condensed matter physics is understanding the electron-scattering processes on the femtosecond time scale. A deep understanding of the electron-phonon \( e-p \) coupling is vital for investigating material properties and improving electronic device technology. The two-temperature model (TTM) is widely used for theoretical analysis \cite{Anisimov1974}, where electrons and phonons are considered independent subsystems with their own temperatures, interacting through a coupling factor \( G_{ep} \). In the original transient TTM, the temporal evolution is expressed as

\begin{equation}
\begin{aligned}
\frac{\partial T_e}{\partial t} &= \frac{\kappa_e}{C_e} \nabla \nabla T_e - \frac{G_{ep}}{C_e} (T_e - T_p), \\
\frac{\partial T_p}{\partial t} &= \frac{\kappa_p}{C_p} \nabla \nabla T_p + \frac{G_{ep}}{C_p} (T_e - T_p).
\end{aligned}
\end{equation}

\begin{table}[h]
    \centering
    \resizebox{\textwidth}{!}{
    \renewcommand{\arraystretch}{1.5}
    \begin{tabular}{|c|c|c|}
        \hline
        \textbf{Carrier} & \textbf{TTM} & \textbf{ETM} \\
        \hline
        \textbf{Electron (e)} & 
        $\frac{\partial T_e}{\partial t} = \frac{\kappa_e}{C_e}\nabla \nabla T_e - \frac{G_{ep}}{C_e} (T_e - T_p)$ 
        & 
        $\tau_e \frac{\partial^2 T_e}{\partial t^2} + \frac{\partial T_e}{\partial t} = \frac{\kappa_e}{C_e}\nabla \nabla T_e $ \\  
        & & $ - \frac{G_{ep,i}^{n}}{C_e} (T_e - T_{p,i}) - \frac{G_{ep,i}^{SC}}{C_e} (T_e - T_{p,i})$ \\
        \hline
        \textbf{Phonon (p)} & 
        $\frac{\partial T_p}{\partial t} = \frac{\kappa_p}{C_p}\nabla \nabla T_p + \frac{G_{ep}}{C_p} (T_e - T_p)$ 
        & 
        $\tau_{p,i} \frac{\partial^2 T_{p,i}}{\partial t^2} + \frac{\partial T_{p,i}}{\partial t} = \frac{\kappa_{p,i}}{C_{p,i}}\nabla \nabla T_{p,i} $ \\  
        & & $ + \frac{G_{ep,i}^{n}}{C_{p,i}} (T_e - T_{p,i}) + \frac{G_{ep,i}^{SC}}{C_{p,i}} (T_e - T_{p,i}) $ \\  
        & & $ + \frac{G_{pp,i}}{C_{p,i}} (T_{lat} - T_{p,i})$ \\
        \hline
    \end{tabular}
    }
    \caption{Comparison of the governing equations of the Two-Temperature Model (TTM) and the Extended Temperature Model (ETM).}
    \label{tab:ttm_etm_carrier}
\end{table}

In the TTM, all phonons are assumed to be in local thermal equilibrium and are assigned to a single average temperature. Here, the TTM is solved on the basis of an avarage thermal conductivity \( k_{p} = 3800\ (W \cdot m^{-1} \cdot K^{-1}) \) and a heat capacity of \( C_{p} = 1.6\ (MJ \cdot m^{-3} \cdot K^{-1}) \) \cite{Cepellotti2015}. This model is based on the assumption that phonons, as collective excitations of the lattice, interact in a way that ensures the phonon distribution reaches equilibrium rapidly compared to other subsystems, such as electrons. The TTM typically focuses on the coupling between two temperatures: the electron temperature and the lattice (phonon) temperature. The approach is often used when the system is not in perfect equilibrium, but the phonons are still assumed to reach a steady-state temperature quickly. In contrast, the ETM recognizes the nonequilibrium nature of the phonon system by distinguishing between different types of phonons, such as acoustic branches, and accounting for the lag effect between these different phonon branches. The ETM introduces a more detailed treatment of phonon transport by considering how the different phonon modes (branches) interact with each other and how the thermal energy is carried between these branches. Lastly, the temperature profiles of the hot electron and {\(pLA\)} are shown in Figure S2. We could easily distinguish three regimes: (1) energy transfer from electron to phonon via
\(e-p\) coupling including the normal and supercollision effect; (2) hot electron cooling and (3) thermal diffusion describing an equilibrium state. 


\begin{figure}[H]
    \centering
    \includegraphics[width=0.8\textwidth, keepaspectratio]{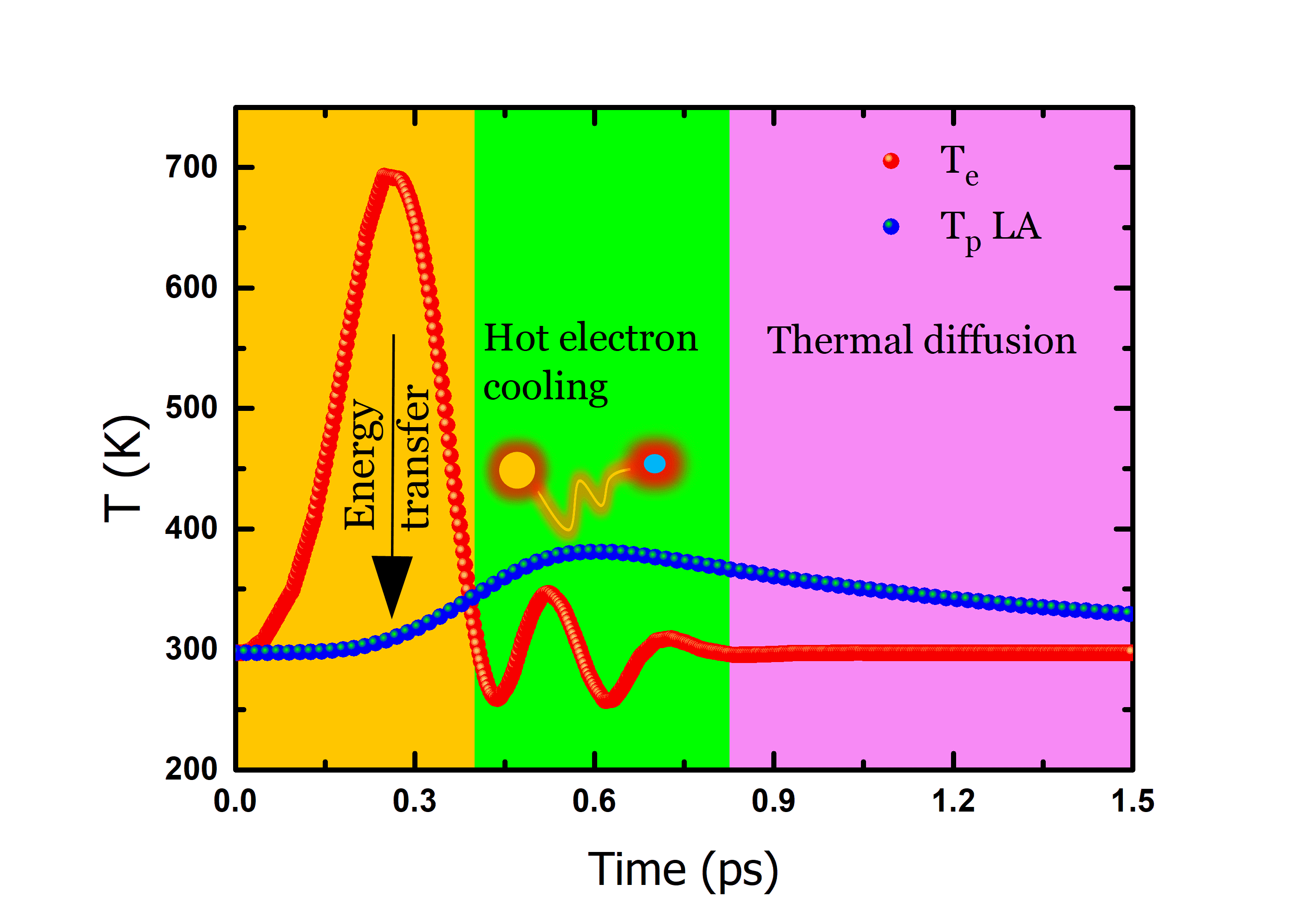}
    \caption{Distribution of the temperature profiles of electron and \(LA\) phonon branch according to the ETM for a pulse width of \(FWHM = 190\,\text{fs}\) at \( Y = 20\,\text{nm} \).}
    \label{fig:fig1}
\end{figure}

To complement Figure 3, we present additional results for intermediate pulse widths of 200 fs, 280 fs, and 320 fs. These simulations provide a more continuous view of how the electron temperature evolves as the excitation pulse duration increases. As shown in Figure S3, the wave-like oscillations in the electron temperature are prominent at shorter pulse widths (e.g., 190 fs) and gradually weaken as the pulse duration increases. This gradual suppression of oscillatory behavior indicates a transition from a wave-dominated to a diffusion-dominated regime of energy relaxation.  To examine the influence of excitation on heat transport, we varied the increase in the peak electronic temperature \(\Delta T_{\text{max}}\). Figure S4 shows the distribution of the electron temperature at \( Y = 20\,\text{nm} \) for a fixed pulse width of 190 fs, with different values of \(\Delta T_{\text{max}}\). As the excitation amplitude increases, the electron temperature exhibits more pronounced wave-like features, indicating stronger non-equilibrium conditions. These effects are attributed to an enhanced \(e-p\) coupling under higher energy deposition, which accelerates energy redistribution and intensifies the cooling dynamics.

\begin{figure}[H]
    \centering
    
    \begin{subfigure}[b]{0.55\textwidth}
        \centering
        \includegraphics[width=\textwidth]{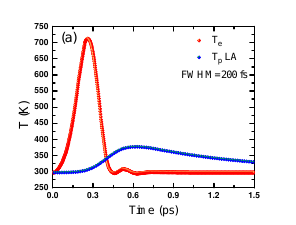}  
        \label{fig:fig2a}
    \end{subfigure}
    \hfill
    \begin{subfigure}[b]{0.55\textwidth}
        \centering
        \includegraphics[width=\textwidth]{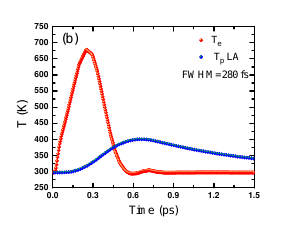}  
        \label{fig:fig2b}
    \end{subfigure}
    
    \vspace{-1em} 
    
    \begin{subfigure}[b]{0.55\textwidth}
        \centering
        \includegraphics[width=\textwidth]{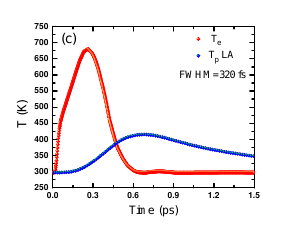}
        \label{fig:fig2c}
    \end{subfigure}
    \hfill
    \caption{Distribution of the temperature profiles of electron and \(LA\) phonon branch according to the ETM at \( Y = 20\,\text{nm} \) for different pulse widths (a) \(FWHM = 200\,\text{fs}\), (b) \(FWHM = 280\,\text{fs}\) and (c) \(FWHM = 380\,\text{fs}\).}
    \label{fig:fig2}
\end{figure}

\begin{figure}[H]
    \centering
    \includegraphics[width=0.8\textwidth, keepaspectratio]{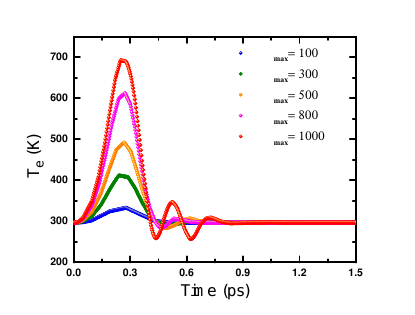}
    \caption{Distribution of the temperature profiles of electron for a pulse width of \(FWHM = 190\,\text{fs}\) at \( Y = 20\,\text{nm} \) for different \(\Delta T_{\text{max}}\). Electron–LA phonon coupling is considered in this analysis.}
    \label{fig:fig1}
\end{figure}